\documentclass[12pt]{article}
\usepackage{graphicx}
\usepackage{amsmath}
\usepackage{amssymb}
\usepackage{caption2}
\begin{document}
\bibliographystyle{prsty}
\begin{center}
{\large {\bf \sc{  Magnetic moment of the pentaquark state $\Theta^+(1540)$  }}} \\[2mm]
Zhi-Gang Wang$^{1}$ \footnote{Corresponding author; E-mail,wangzgyiti@yahoo.com.cn.  },  and Ri-Cong Hu$^{2} $    \\
$^{1}$ Department of Physics, North China Electric Power University, Baoding 071003, P. R. China \\
$^{2}$ Department of Mathematics, North China Electric Power University, Baoding 071003, P. R. China \\
\end{center}

\begin{abstract}
In this article, we study the magnetic moment of the pentaquark $
\Theta^+(1540)$ as (scalar)diquark-(pseudoscalar)diquark-antiquark
state with the QCD sum rules approach in the external
electromagnetic field. Due to the special structure of the
interpolating current, only the electromagnetic interactions of the
$s$ quark  with the external  field have contributions to the
magnetic moment with  tensor structure $\{\sigma_{\mu\nu} {\hat p}
+{\hat p}\sigma_{\mu\nu}\}$.  The numerical results indicate the
magnetic moment of the pentaquark state $ \Theta^+(1540)$ is about
$\mu_{\Theta^+}=(0.16\pm0.03)\mu_N$.
\end{abstract}


\section{Introduction}

Intense theoretical investigations have been motivated
 to clarify the quantum numbers and to understand the
under-structures of the pentaquark state $\Theta^+(1540)$ since its
observation \cite{exp2003,DiakonovPentaquark,ReviewPenta}. The
extremely narrow width  below $10MeV$ puts forward a serious
challenge to all theoretical models, in the conventional
uncorrelated quark models  the expected width is supposed to be of
the order of several hundred $MeV$, since the strong decay
$\Theta^{+} \, \rightarrow K^+ N$ is Okubo-Zweig-Iizuka (OZI)
super-allowed. The zero of the third component of isospin $I_3=0$
and the absence of isospin partners suggest  that the baryon
$\Theta^+(1540)$ is an isosinglet, while the spin and parity have
not been experimentally determined yet and  no consensus has ever
been reached  on the theoretical side.
 Determining the parity of the pentaquark state  $\Theta^+(1540)$
 is of great importance in establishing its basic quantum
numbers and in understanding the low energy QCD especially when
multiquarks are involved.
The experiments of photo- or electro-production and proton-proton collision can be used
  to determine the fundamental quantum numbers of  the pentaquark state $\Theta^+(1540)$, such
  as spin and parity \cite{HosakaPP}.
  The magnetic moment of the $\Theta^+(1540)$ $\mu_{\Theta^+}$ is an important ingredient
in studying the cross sections of the photo-production,  and may be
extracted from the experiments eventually in the  future. In fact,
the magnetic moments of the pentaquark states are
  fundamental parameters as their  masses, which have  copious  information  about the underlying quark
structures, can be used to
distinguish the preferred quark configurations from  various  theoretical models and  deepen our
understanding of the underlying dynamics.

 There have been several works on the magnetic moment $\mu_{\Theta^+}$
  \cite{MagneticMoment,Huang04M,WangYW,WangWY},
in this article, we take the point of view that the quantum
numbers of the pentaquark state  $\Theta^{+}(1540)$ are $J=\frac{1}{2}$ ,
$I=0$ , $S=+1$, and study its magnetic moment $\mu_{\Theta^+}$ with the QCD sum rules approach \cite{Shifman79}.

 The article is arranged as follows:   we derive the QCD sum rules in the external
 electromagnetic field for the magnetic moment $\mu_{\Theta^+}$ in section II;
 in section III, numerical results; section IV is
reserved for conclusion.

\section{QCD Sum Rules in External Electromagnetic Field}
In the following, we write down  the two-point correlation function
$\Pi_{\Theta}(p)$ in the presence of a weak external electromagnetic
field $F_{\alpha\beta}$ \cite{Ioffe84}\footnote{For technical
details, one can consult Refs.\cite{WangYW,WangWY}. },
\begin{eqnarray}
\Pi_{\Theta}(p)&=& i\int d^4 x e^{ip\cdot x}\langle 0|T \{ J(x){\bar
J}(0) \}|0\rangle_{F_{\alpha\beta
} } \, , \nonumber \\
&=&\Pi_0(p) + \Pi_{\mu\nu} (p) F^{\mu\nu} +\cdots \, ,
\end{eqnarray}
where the $\Pi_0(p)$ is the correlation function  without the
external field $F_{\alpha\beta}$ and the $\Pi_{\mu\nu} (p)$ is the
linear response term. In this article, we take the
(scalar)diquark-(pseudoscalar)diquark-antiquark type interpolating
current $J(x)$ for the pentaquark state $\Theta^+(1540)$
\cite{Sugiyama04},
\begin{eqnarray}
  J(x)&=&\epsilon^{abc}\epsilon^{def}\epsilon^{cfg}
   \{u_a^T(x)Cd_b(x)\}\{u_d^T(x)C\gamma_5 d_e(x)\}C\bar{s}_g^T(x) ,
  \end{eqnarray}
here the $a$, $b$, $c$ $\cdots$ are color indexes. There have been
several works on the magnetic moment of the $\Theta^+(1540)$ using
the diquark-triquark type current $J_1(x)$ and
diquark-diquark-antiquark type current $J_2(x)$
  within the framework of the QCD sum rules approach
\cite{Huang04M,WangYW,WangWY},
\begin{eqnarray}
J_1(x)&=&{1\over \sqrt{2}} \epsilon^{abc} \left\{u^T_a(x) C\gamma_5
d_b (x)\right\} \{ u_e (x) {\bar s}_e (x) i\gamma_5 d_c(x) - d_e (x)
{\bar s}_e (x) i\gamma_5 u_c(x)  \} \, ,\\
 J_2(x)&=&\left\{t \eta_1(x) + \eta_2(x) \right\} , \\
\eta_1(x)&=&{1\over\sqrt{2}}\epsilon^{abc}\left\{\left[u_a^T(x) C
\gamma_5 d_b(x)\right] \left[u_c^T(x) C \gamma_5
d_e(x)\right]C\bar{s}^T_e(x) - (u\leftrightarrow d)\right\} , \nonumber\\
\eta_2(x)&=&{1\over\sqrt{2}}\epsilon^{abc}\left\{\left[u_a^T(x) C
d_b(x)\right] \left[u_c^T(x) C  d_e(x) \right ]C\bar{s}^T_e(x) -
(u\leftrightarrow d)\right\} , \nonumber
\end{eqnarray}
There exist a great number of  possible quark configurations satisfy
the Fermi statistics and the color singlet condition
 for the under-structures of the pentaquark state $\Theta^+(1540)$,
  if we release  stringent dynamical constraints.
  Different quark configurations can be implemented by different
interpolating currents, which are always  lead to substantially
different magnetic moments. Although the  three interpolating baryon
currents $J$, $J_1(x)$, $J_2(x)$ can all give satisfactory masses
about $1.5GeV$ for the $\Theta^+(1540)$, the resulting magnetic
moments are  substantially different. Without additionally  powerful
constraints (such as the magnetic moment), we can not select the
preferred quark configurations from various theoretical quark
models.

The Fierz re-ordering of the interpolating current $J(x)$  can lead
to the following sub-structures,
\begin{eqnarray}
    J(x) &=&{1\over 4}\epsilon^{abc}(u^{T}_a C d_b)\left\{-d_c(\bar s\gamma_5 u)
  -\gamma^\mu d_c(\bar s\gamma_5\gamma_\mu u)\right.\cr
  &&\left.+\frac{1}{2}\sigma^{\mu\nu}d_c(\bar s\sigma_{\mu\nu}\gamma_5 u)
  +\gamma^\mu\gamma_5 d_c(\bar s\gamma_\mu u)
  -\gamma_5 d_c(\bar su)-(u \leftrightarrow d)\right\}.
\end{eqnarray}
A naive result  of the Fierz re-ordering may be the appearance of
the reducible contributions with the sub-structure of $udd-u\bar{s}$
(i.e. $N-K$) clusters   in the two-point correlation function
$\Pi_0(p)$ \cite{Reducible}, however, in our calculations with the
interpolating current $J(x)$ in Eq.(2), no such factorable
$udd-u\bar{s}$ terms appear, so there are no reducible $N-K$
contributions to the correlation function $\Pi_0(p)$.  The
re-ordering in Dirac spin space is always  accompanied  with  color
re-arrangement, which involves the underlying dynamics.  The
appearance of the $N-K$ component in the Fierz re-ordering maybe
manifest the possibility (not the probability) of the evolution from
the $\Theta^+(1540)$ to the $NK$ final state without net
quark-antiquark pairs  creation (maybe the quark-antiquark pairs
created and annihilated subsequently), which is significantly in
contrast to the conventional baryons, however, we have no knowledge
about the detailed process of the evolution.  If there are really
some $N-K$ components in the interpolating current $J(x)$, they
should be factorized out, the remainder can not have the correct
quantum numbers to interpolate  the $\Theta^+(1540)$. For detailed
discussions about this subject, one can see Ref.\cite{WangYW05} .

 The interpolating
current $J(x)$ in Eq.(2) can couple to the pentaquark states  with
both negative and positive parity, and picks out only the state with
the lowest mass without knowledge about its parity, for example, we
use the Ioffe current $J_{p}(x)=\epsilon^{abc} ({u}^{T}_a(x)C\gamma
_{\mu}u_{b}(x)) \gamma _5 \gamma ^{\mu}d_c(x)$ to interpolate the
 proton and pick out only the lowest energy  state \cite{Ioffe81}, which happen
 to have positive parity,  while the possible
negative parity states are included in the high resonances and
continuum states. It is not necessary to include both the negative
and positive states in the phenomenological spectral density.  As
the electromagnetic vertex of the pentaquark states with either
negative or positive parity is the same, we can extract the absolute
value of the magnetic moment of the lowest pentaquark state.

The linear response term $\Pi_{\mu\nu}(p)$ in the weak external
electromagnetic field $F_{\alpha\beta}$  has three different Dirac
tensor structures,
\begin{equation}
\Pi_{\mu\nu}(p) =\Pi (p) \left\{\sigma_{\mu\nu} {\hat p} +{\hat
p}\sigma_{\mu\nu}\right\} +\Pi_1(p)
i\left\{p_{\mu}\gamma_{\nu}-p_{\nu}\gamma_{\mu}\right\}{\hat p}
+\Pi_2(p) \sigma_{\mu\nu}  \, .
\end{equation}
The first structure has an odd number of $\gamma$-matrix and conserves chirality, the second and third have even
number of $\gamma$-matrixes and violate chirality. In the original QCD sum rules analysis of the nucleon magnetic
 moments \cite{Ioffe84}, the interval of dimensions (of the condensates) for the odd
structure is larger than the interval of dimensions for the even structures, one may expect a better accuracy of the results
obtained from the sum rules at the odd structure. In this article, the spin of the pentaquark state $\Theta^+(1540)$ is supposed
to be $\frac{1}{2}$, just like the nucleon,
we can choose  the first Dirac tensor structure $\left\{\sigma_{\mu\nu} {\hat p} +{\hat p}\sigma_{\mu\nu}\right\}$.
The phenomenological  spectral density can be written as
\begin{equation}
\frac{\mbox{Im} \Pi (s)}{\pi} = {1\over 4} \{F_1(0)+F_2(0)\} f_0^2
\delta^\prime (s-m_{\Theta^+}^2) + C_{subtract} \delta
(s-m_{\Theta^+}^2) +\cdots\, ,
\end{equation}
where the first term corresponds to the magnetic moment $\mu_{\Theta^+}$, and is
of double-pole. The second term comes from the electromagnetic transitions between the pentaquark state $\Theta^+(1540)$
 and the excited states,  and is of single-pole.
Here we introduce the quantity  $C_{subtract}$ to parameterize
 the electromagnetic transitions between the ground pentaquark state and the high
 resonances, it may have   complex dependence  on the energy $s$ and high
 resonance masses. However, we have no knowledge about the high resonances,
 even the existence  of the ground pentaquark state $\Theta^+(1540)$
 is not firmly established, which is in contrast to the conventional baryons, in those channels we can use
 the experimental data as guides  in constructing the phenomenological spectral densities.
 In practical manipulations, we can take the $C_{subtract}$ as
 an  unknown constant, and fitted to reproduce reliable values for
 the form factors $F_1(0)+F_2(0)$.
 The higher
resonances and  continuum states in Eq.(7) are neglected  for
simplicity.
 From the electromagnetic form factors  $F_1(0)$ and $ F_2(0)$ , we can obtain the
 magnetic moment  $\mu_{\Theta^+}$,
\begin{equation}
\mu_{\Theta^+}=\left\{ F_1(0)+F_2(0)\right\} \frac{e_{\Theta^+}}{2m_{\Theta^+}}.
\end{equation}
After performing the  operator product expansion in the  deep Euclidean space-time region,
we can express the correlation functions at the level of quark-gluon
degrees of freedom into the following form through dispersion
relation,
  \begin{eqnarray}
  \Pi(P^2)= \frac{e_s}{\pi}\int_{m_s^2}^{s_0}ds
  \frac{{\rm Im}[A(s)]}{s+P^2}+\cdots\, ,
  \end{eqnarray}
where
\begin{eqnarray}
\frac{{\rm Im}[A(s)]}{\pi}&=& \frac{s^4}{2^{12}5!4!\pi^8}-\frac{
m_s\chi \langle\bar{s}s \rangle s^3}{2^{7}5!4!\pi^6}
+\frac{s^2}{2^{13} 4!\pi^6} \langle \frac{\alpha_s GG}{\pi}\rangle .
\end{eqnarray}
The presence of the external electromagnetic field $F_{\mu\nu}$ induces
three new vacuum condensates i.e. the vacuum susceptibilities $\chi$, $\kappa$ and $\xi$ in the QCD vacuum
   \cite{Ioffe84}.  The values with  different theoretical
approaches are different from each other, for a short review,  one
can see Ref.\cite{Wang02}. Here we shall adopt the values
$\chi=-4.4\, \mbox{GeV}^{-2}$, $\kappa =0.4$ and $\xi = -0.8$
\cite{Ioffe84,Belyaev84}. From Eqs.(9-10), we can see that due to
the special structure of the diquark-diquark-antiquark type
interpolating current $J(x)$ (  also the $J_2(x)$ \cite{WangWY}),
the $u$ and $d$ quarks which constitute the diquarks have no
contributions to the magnetic moment
 though they have electromagnetic interactions with the external
field, the net contributions to the magnetic moment come from the
$s$ quark only, which is  different significantly from the results
obtained in Refs.\cite{Huang04M,WangYW} with the diquark-triquark
type interpolating current $J_1(x)$. Although the
diquark-diquark-antiquark type and diquark-triquark type
configurations implemented by the interpolating currents $J(x)$ ( or
$J_1(x)$ ) and $J_2(x)$ respectively can give satisfactory masses
for the pentaquark state $\Theta^+(1540)$,  the resulting magnetic
moments are substantially different, once the magnetic moment can be
extracted from the electro- or photo-production experiments, we can
select the preferred  configuration.

 Finally we obtain the sum rules for the form factors   $F_1(0)$ and $F_2(0)$,
\begin{eqnarray}
-\frac{1}{4} \left\{F_1(0)+F_2(0)\right\}\frac{1+CM^2}{M^4}f_0^2e^{-\frac{m^2_{\Theta^+}}{M^2}}=\frac{1}{M^2}\int_{m_s^2}^{s_0}
ds  \frac{{\rm Im}[A(s)]}{\pi}e^{-\frac{s}{M^2}},
\end{eqnarray}
where the definition  $\langle 0| J(0) |\Theta^+ (p)\rangle =f_0
u(p)$ has been used, the $m_s$ is the strange quark mass and $s_0$
is the threshold parameter used to subtract the contributions from
the higher resonances and  continuum states. The Borel transform can
not eliminate the contaminations  from the single-pole terms, we
introduce
 the parameter  $C$ which proportional to the  $C_{subtract}$ in Eq.(7) to the subtract the contaminations.
  We have no knowledge about the  electromagnetic transitions between the pentaquark state $\Theta^+(1540)$
 and the excited states (or high resonances), the $C$ can be taken as a free parameter, we  choose the suitable values for $C$ to
  eliminate the contaminations from  the single-pole terms to obtain the reliable  sum rules. The contributions from
  the single-pole terms may as large as or larger than the double-pole term, in practical calculations,
   the $C$ can be fitted to give stable sum rules with respect to
 variations  of the Borel parameter $M^2$ in a suitable interval. Taking  the $C$ as an  unknown constant has
 smeared the complex energy $s$ and high resonances masses dependence, which will certainly impair the
 prediction
 power. As  there really exists a platform with the variations  of the Borel parameter $M^2$, the
 predictions still make sense.
 Furthermore, from the correlation function $\Pi_0(p)$ in Eq.(1), we can
 obtain the sum rules for the coupling constant $f_0$ \cite{Sugiyama04},
\begin{equation}
f^2_0e^{-{m_{\Theta^+}^2\over M^2}}=\int_{m_s^2}^{s_0} ds e^{-{s\over M^2}}
\rho_0 (s),
\end{equation}
where
\begin{eqnarray}
 \rho_0=\frac{s^5}{2^{10}5!5!7\pi^8}+\frac{m_s \langle
\bar{s}s\rangle s^3}{2^{8}5!3!\pi^6}-\frac{m_s \langle
\bar{s}g_s \sigma  G s\rangle
s^2}{2^{9}4!3!\pi^6}+
\frac{s^3}{2^{10}5!3!\pi^6}\langle \frac{\alpha_s
GG}{\pi}\rangle . \nonumber
\end{eqnarray}

\section{Numerical Results}
The input parameters are taken as   $\chi=-(4.4\pm0.4)GeV^{-2}$,
 $\langle \bar{s}s \rangle=(0.8\pm0.1)\langle
\bar{q}q \rangle$, $\langle \bar{s}g_s\sigma  G s
\rangle=m_0^2\langle \bar{s}s \rangle$, $m_0^2=(0.8\pm0.1)GeV^2$,
$\langle \bar{q}q \rangle=-(0.24\pm0.01 GeV)^3$, $\langle
\frac{\alpha_sGG}{\pi} \rangle=(0.33GeV)^4$, $m_u=m_d=0$ and
$m_s=(140\pm10)MeV$.
 Here we use the standard values \cite{Shifman79},  small variations of those condensates will not
 lead to large changes about the numerical
 values. The threshold parameter $\sqrt{s_0}$  is chosen to  vary between $(1.7-1.9) GeV$ to avoid possible contaminations from
  higher resonances and continuum states. In the region $M^2=(1.5-3.5)GeV^2$, the sum rules for $F_1(0)+F_2(0)$ are almost independent of
 the Borel parameter $M^2$, which are shown in Fig.1 for  $\chi=-4.4GeV^{-2}$,
 $\langle \bar{s}s \rangle=0.8\langle
\bar{q}q \rangle$,  $m_0^2=0.8GeV^2$, $\langle \bar{q}q
\rangle=(-0.24 GeV)^3$, $m_s=140MeV$.
\begin{figure}
 \centering
 \includegraphics[totalheight=7cm]{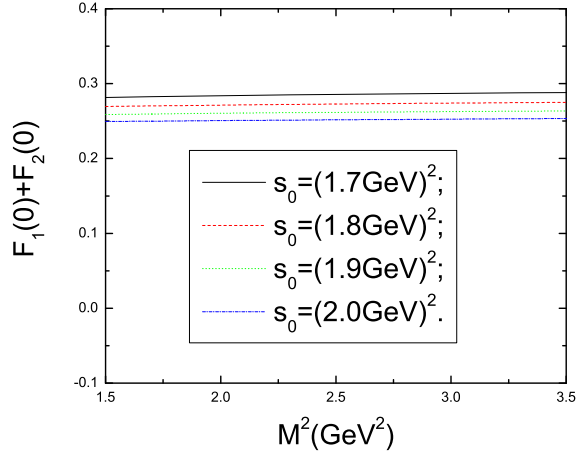}
 \caption{$|F_1(0)+F_2(0)|$ with the  Borel Parameter $M^2$. }
\end{figure}
 For $\sqrt{s_0}=(1.7-1.9) GeV$, we obtain the values
 \begin{eqnarray}
 F_1(0)+F_2(0)&=&0.27\pm 0.05 \, , \nonumber\\
 \mu_{\Theta^+}&=& (0.27\pm 0.05) \frac{e_{\Theta^+}}{2m_{\Theta^+}}\, , \nonumber \\
  &=& (0.16\pm0.03)\mu_N,
  \end{eqnarray}
where the $\mu_N$ is the nucleon  magneton.
 Although  the numerical values for
 the magnetic moment  $ \mu_{\Theta^+}$  vary with theoretical
 approaches (i.e. $ \mu_{\Theta^+}\approx (0.1-0.7)\mu_N$ or $ \mu_{\Theta^+}\approx- (0.1-1.2)\mu_N$), they are small in general;  our numerical results
are consistent with most of the existing values of theoretical
estimations \cite{MagneticMoment,Huang04M,WangYW,WangWY}.  For a
short review of the existing calculations of the magnetic moments of
the $\Theta^+(1540)$, one can consult Ref.\cite{WangYW}. The main
contributions to the magnetic moments $\mu_{\Theta^+}$ come from the
perturbative term in Eq.(10), about $70\%$, the contributions from
terms of the quark condensates and gluons condensates   have
opposite sign,  and the resulting net contributions are about
$30\%$, the high dimensional condensates are neglected as they are
suppressed by large denominators. For the conventional ground state
mesons and baryons, due to the resonance dominates over the QCD
continuum contributions, the good convergence of the operator
product expansion, and the useful experimental guidance on the
threshold parameter $s_0$, we can obtain the fiducial Borel mass
region. However, in the QCD sum rules for the pentaquark states, the
spectral density $\rho(s) \sim s^m$ with $m$ larger than  the
corresponding ones in the
 sum rules for the conventional baryons, larger $m$ means   stronger dependence on the
continuum or the threshold parameter $s_0$ \cite{Narison04}. In
Eq.(12), due to the large  continuum contributions,  the threshold
parameter  $s_0$ has  to be fixed ad hoc or intuitively. In this
article, the threshold parameter $s_0$ are taken to be
$\sqrt{s_0}=(1.7-1.9)GeV$, the mass $m_{\Theta^+}=1540MeV$ and the
width $\Gamma_{\Theta}<10MeV$, the contributions from the lowest
pentaquark state can be successfully included in.
 Although
the uncertainties of the condensates, the neglect of the higher
dimension condensates, the lack of perturbative QCD corrections,
etc, will  result in errors, we have stable sum rules, which are
shown in Fig.1, the predictions still make sense, or qualitative at
least.

In Ref.\cite{WangWY}, the authors take the diquark-diquark-antiquark
current $J_2(x)$ which is linear superposition of both S-type and
P-type baryon currents ( $\eta_1(x)$ and $\eta_2(x)$) to calculate
the magnetic moment $\mu_{\Theta^+}$ in two approaches (i.e. the QCD
sum rules in the external field and the light-cone QCD sum rules),
and obtain $\mu_{\Theta^+}=-(0.11\pm0.02)\mu_N$ and
$\mu_{\Theta^+}=-(0.1\sim0.5)\mu_N$, respectively. As the values
obtained from the QCD sum rules in the external field are more
stable than the corresponding ones from the light-cone QCD sum
rules, $\mu_{\Theta^+}=-(0.11\pm 0.02)\mu_N$ is more reliable. For
the diquark-diquark-antiquark type interpolating currents $J(x)$ and
$J_2(x)$, only the electromagnetic interactions of  the $s$ quark
with the external field  have contributions to magnetic moment with
the tensor structure $\{\sigma_{\mu\nu} {\hat p} +{\hat
p}\sigma_{\mu\nu}\}$, the resulting magnetic moments are
substantially different. For the diquark-triquark type interpolating
currents $J_1(x)$, the electromagnetic interactions of all the $u$,
$d$ and $s$ quarks with the external field  have contributions to
the magnetic moment, and $\mu_{\Theta^+}=(0.24\pm0.02)\mu_N$
\cite{WangYW}. The interpolating currents $J(x)$, $J_1(x)$ and
$J_2(x)$ with different quark configurations can all give
satisfactory masses for the $\Theta^+(1540)$, without additional
powerful dynamical constraints, we can not pick  out the preferred
configurations from various theoretical models ( constituent quark
models  or cluster quark models ) .  The magnetic moment plays  an
important role in understanding the under-structures of the
pentaquark state.

\section{Conclusion }

In summary, we have calculated  the magnetic moment of the
 pentaquark  $\Theta^+(1540)$ as (scalar)diquark-(pseudoscalar)diquark-antiquark state with the  QCD sum
rules  approach in the weak external electromagnetic field. The
numerical results are consistent with most of the existing values of
theoretical estimations, $ \mu_{\Theta^+}= (0.27\pm 0.05)
\frac{e_{\Theta^+}}{2m_{\Theta^+}}
  = (0.16\pm0.03)\mu_N$ .
The  magnetic moments of the
baryons are  fundamental parameters as their masses, which have
copious  information  about the underlying quark structures,
different substructures can lead to
 very different results. The  small magnetic moment $\mu_{\Theta^+}$ may be
 extracted from the electro- or photo-production
experiments eventually in the future, which may be used to
distinguish the preferred quark configurations and QCD sum rules
from various theoretical models, obtain more insight into the
relevant degrees of freedom and deepen our understanding about  the
underlying dynamics that determines the properties of the exotic
pentaquark states.

\section*{Acknowledgment}
This  work is supported by National Natural Science Foundation,
Grant Number 10405009,  and Key Program Foundation of NCEPU. The
authors are indebted to Dr. J.He (IHEP), Dr. X.B.Huang (PKU) and Dr. L.Li (GSCAS)
for numerous help, without them, the work would not be finished.

\end{document}